\documentclass{JHEP3}
\pdfoutput=1

\usepackage{amsmath}
\usepackage{amssymb}
\usepackage{graphicx}
\usepackage{slashed}

\newcommand*{\longhookrightarrow}{\ensuremath{\lhook\joinrel\relbar\joinrel\rightarrow}}
\newcommand*{\longtwoheadedrightarrow}{\ensuremath{\relbar\joinrel\twoheadrightarrow}}
\newcommand{\field}[1]{\mathbb{#1}}

\newcommand{\bC}{\mathbb{C}}
\newcommand{\bF}{\mathbb{F}}
\newcommand{\bP}{\mathbb{P}}
\newcommand{\bR}{\mathbb{R}}
\newcommand{\bZ}{\mathbb{Z}}
\newcommand{\cA}{\mathcal{A}}
\newcommand{\cD}{\mathcal{D}}
\newcommand{\cF}{\mathcal{F}}
\newcommand{\cL}{\mathcal{L}}

\newcommand{\cN}{\mathcal{N}}
\newcommand{\cO}{\mathcal{O}}
\newcommand{\cU}{\mathcal{U}}

\title{Global F-theory Models: Instantons and Gauge Dynamics}

\author{Mirjam Cveti\v{c}, I\~naki Garc\'ia-Etxebarria and James Halverson\\
  Department of Physics and Astronomy, University of Pennsylvania,\\
  Philadelphia, PA 19104-6396, USA\\
  and\\
  Kavli Institute for Theoretical Physics, Kohn Hall,\\
  UCSB, Santa Barbara, CA 93106, USA\\
  E-mail: \email{cvetic@cvetic.hep.upenn.edu},
  \email{inaki@sas.upenn.edu}, \email{jhal@physics.upenn.edu}}

\abstract{We elucidate certain aspects of F-theory gauge dynamics, due
  to quantum splitting of certain brane stacks, which are absent in
  the Type IIB limit. We also provide a working implementation of an
  algorithm for computing cohomology of line bundles on arbitrary
  toric varieties. This should be of general use for studying the
  physics of global Type IIB and F-theory models, in particular for
  the explicit counting of zero modes for rigid F-theory instantons
  which contribute to charged matter couplings. We illustrate the
  discussion by constructing and analyzing in detail a compact
  F-theory GUT model in which a D-brane instanton generates the top
  Yukawa coupling non-perturbatively.}

\preprint{UPR-1216-T\\NSF-KITP-10-035}

\begin{document}

\section{Introduction}

F-theory \cite{Vafa:1996xn} provides a promising framework for studies
of string vacua with potentially realistic particle
physics.\footnote{For recent efforts within the Type IIA intersecting
  D-brane framework, see \cite{Blumenhagen:2005mu,Blumenhagen:2006ci}
  for a review and \cite{Cvetic:2009yh, Cvetic:2009ez, Cvetic:2009ng,
    Cvetic:2010mm} for a systematic study of local realistic MSSM
  quivers.}  It combines many of the nice features of type IIB,
particularly localization of gauge degrees of freedom on D-branes,
with some of the nice features of heterotic models, such as the
natural appearance of exceptional groups.

Most effort so far has centered on the classical aspects of F-theory
models, i.e. couplings that can be computed as wave function
overlaps. This perturbative sector is already enough for constructing
appealing and phenomenologically promising models (there is a rapidly
growing literature on the topic, starting with
\cite{Donagi:2008ca,Beasley:2008dc,Beasley:2008kw,Donagi:2008kj}).
Nevertheless, non-perturbative corrections to this picture can in some
instances be the dominant contribution, and modify the picture
substantially.

Classical examples come from gaugino condensation or euclidean
instantons wrapping isolated cycles. Under favorable conditions
\cite{Witten:1996bn}, they can generate a superpotential for the
K\"ahler modulus associated with the cycle. This effect has very
important applications for moduli stabilization in IIB and F-theory
\cite{Balasubramanian:2005zx,Denef:2004dm}, and it is an essential
ingredient of many semi-realistic type IIB scenarios
\cite{Kachru:2003aw}.

Another important effect coming from D-brane instantons has been
greatly clarified in the type II context in recent years. Whenever a
D-brane instanton intersects a D-brane stack, there are some zero
modes in the instanton worldvolume that are charged under the gauge
symmetry on the D-brane stack. Integration over these charged zero
modes can generate F-term couplings for matter fields living on the
D-brane stack
\cite{Ganor:1998ai,Blumenhagen:2006xt,Ibanez:2006da,Florea:2006si}.

In the type II context these charged instantons solve a long-standing
difficulty in constructing realistic models: brane stacks have $U(1)$
factors that survive perturbatively, and generally forbid certain
important couplings in the MSSM lagrangian. Typical examples are the
top-quark Yukawa couplings in $SU(5)$ GUT models, and the $\mu$
term. Charged D-brane instantons do not necessarily respect
perturbative $U(1)$ symmetries, and thus can generate these couplings
(see \cite{Blumenhagen:2009qh} for a recent review).

A very attractive feature of F-theory compactifications is that, due
to their close relation with exceptional groups, these $U(1)$ factors
can be absent, and thus the couplings which are problematic from the
type II point of view can be obtained perturbatively. Nevertheless,
one may still investigate the effect of non-perturbative effects in
F-theory. The motivations are many: instanton effects can be naturally
suppressed, depending on the volume of the cycle wrapped by the
instanton, a feature that can be quite convenient whenever one desires
to obtain a hierarchy. One may also want to try a hybrid approach,
building a good model in the better understood IIB context with some
couplings coming from euclidean instantons and then uplifting to
F-theory to improve some aspects of the model. Finally, and perhaps
most importantly, non-perturbative effects will be there in any case,
and one must be able to understand how they affect the model at hand.

With this motivation in mind, in this paper we discuss in detail the
F-theory uplift of a particular global type IIB model in which D-brane
instantons are known to play a crucial role in generating MSSM
couplings, in particular the top-quark Yukawa coupling. We use this
model as a prototype illustrating the two major themes of this paper:
the quantum splitting of classical brane stacks, and our
implementation of line bundle cohomology computations using the \v
Cech complex. These are issues that are important for any realistic
F-theoretical model building, and we are explicit about their
resolution. Before going into a complete discussion, let us briefly
review both issues in turn.

The first issue, quantum brane splitting, arises when uplifting
interesting weakly coupled IIB models to F-theory. It can easily
happen, as in our particular example, that brane stacks that make
perfect sense in IIB split into sub-stacks when uplifted to
F-theory. This is due to non-perturbative D$(-1)$ instantons that
start modifying the geometry as soon as $g_s$ is non-vanishing. A
convenient way of studying this problem is using D3 brane probes, for
which the D$(-1)$ effects appear as ordinary gauge instantons, and the
ambient geometry appears as the geometry of the Coulomb branch of
moduli space. The effects of D$(-1)$ instantons on the geometry can
thus be understood using Seiberg-Witten theory. We do this in
section~\ref{sec:quantum-splitting}, explaining in which cases the
quantum modification occurs, and in which it does not.

Line bundle cohomology computations arise naturally when analyzing
neutral instanton zero modes, which in F-theory are counted by the
cohomology groups of the trivial sheaf on the worldvolume of the
instanton \cite{Witten:1996bn}. The framework in which we work is
toric geometry\footnote{We recommend \cite{Skarke:1998yk,
    Closset:2009sv,Bouchard:2007ik} for an introduction and
  \cite{Cox:LecturesOnToricVarieties, CoxLittleSchenck:ToricVarieties,
    Fulton} for a more thorough treatment.}, and accordingly our
Calabi-Yau fourfold $Y$ will be a complete intersection in a six
(complex) dimensional ambient space $X_{\Sigma^{''}}$, specified by
toric data, with instantons wrapping divisors of $Y$. Sheaf cohomology
on the instanton can then be computed by first computing the
cohomology of the toric \v Cech complex (twisted by $\cL$) on
$X_{\Sigma^{''}}$, and then using the Koszul complex to project down
results to divisors of $Y$. We have collected and reviewed the
relevant mathematical background in section~\ref{sec:cech}. The
algorithm for computing \v Cech cohomology that we review there, while
straightforward, quickly becomes intractable if done by hand. Luckily,
it is not hard to instruct a computer to do it, and we provide a
working implementation that should be useful for doing general
computations of line bundle cohomology on arbitrary toric varieties.

\medskip

Previous works on charged instantons in F-theory include
\cite{Blumenhagen:2010ja}, which lays down part of the framework
required for studying charged instantons, and
\cite{Heckman:2008es,Marsano:2008py}, which propose to use euclidean
D-branes to implement local F-theory models for GMSB (although one has
to be careful in determining which instantons can be responsible for
supersymmetry breaking \cite{Cvetic:2009mt,Cvetic:2009ah}).

\medskip

This paper is organized as follows. In
section~\ref{sec:quantum-splitting} we analyze the quantum splitting
of classical brane stacks, determining in which cases it occurs. In
section~\ref{sec:cech} we discuss the mathematical background required
for computing zero modes of instantons in the toric context, and
provide the link to our implementation of the algorithm for computing
\v Cech line bundle cohomology on toric
varieties. In section~\ref{sec:example} we illustrate these general
considerations in a particular example with interesting
phenomenological features.

\medskip

While we were writing our results, we became aware of the work
\cite{Blumenhagen:2010pv}, which provides an efficient algorithm and
computer implementation for computing line bundle cohomology on toric
varieties, and thus overlaps with our discussion in
section~\ref{sec:cech}. We thank the authors of that work for sharing
their insights.

\section{F-theory Gauge Dynamics}
\label{sec:quantum-splitting}

Suppose that we are given a type IIB compactification with a stack of
D7 branes. We are interested on lifting this configuration to
F-theory, and in particular on the fate of the stack of D7 branes. We
will see that in some cases the brane stack splits, and the
F-theoretical picture is qualitatively different. For concreteness, in
most of the analysis we will take the gauge group on the stack to be
$SO(6)$, as that is the one that will make an appearance in our
example in section~\ref{sec:example}, but the discussion generalizes
easily and we give some related results at the end of our analysis.

\subsection{Description from Seiberg-Witten Theory}

The basic physics at play here is similar to that which smooths out
the $O7^-$ plane in F-theory \cite{Sen:1997gv,Sen:1997kw}. It can be
elucidated by studying the world-volume dynamics of a $D3$ brane probe
close to the $SO(6)$ stack \cite{Banks:1996nj}. The theory on the
worldvolume of the D3 brane has a Coulomb branch, with a Coulomb
branch parameter that can be identified with the position of the D3
brane in the direction transverse to the $SO(6)$ stack. Furthermore,
the exact solution of the gauge theory on the probe can be described
in terms of an elliptic fibration over the Coulomb branch (the
Seiberg-Witten solution \cite{Seiberg:1994rs,Seiberg:1994aj}), which
can be identified with the F-theory geometry in which the probe moves
\cite{Sen:1997gv,Sen:1997kw,Banks:1996nj}.

We construct the $SO(6)$ theory by putting three D7 branes on top of
an orientifold. Due to the orientifold, the worldvolume theory on the
D3 brane probe is an $\cN=2$ $SU(2)$ theory, and due to the three $A$
branes we have 3 massless quarks in the theory (at weak coupling).
The Seiberg-Witten curve for the theory with three massless flavors is
given by \cite{Seiberg:1994aj}:
\begin{equation}
  \label{eq:SW-massless}
  y^2 = x^2(x-u) + t\Lambda_3^2(x-u)^2
\end{equation}
with $u$ the coordinate on the Coulomb branch in the $SU(2)$ theory,
$\Lambda_3^2$ the strong coupling scale of the theory, and $t$ a
constant which we could absorb in the definition of $\Lambda_3^2$. The
elliptic fibration~\eqref{eq:SW-massless} over the complex $u$ plane
degenerates over two points. There is a degeneration of order one at
$u=-t\Lambda_3^2/4$, and a degeneration of order 4 at $u=0$. This
already explains why we could not obtain the $SO(6)$ singularity
above: the coupling $\Lambda_3$ of the $SU(2)$ theory is determined by
the string coupling at the position of the D3. In the perturbative IIB
limit this coupling is everywhere vanishing, so $\Lambda_3\to 0$, and
the two singular points collide, enhancing the degeneration to order
5, as we expected from Tate's classification. Nevertheless, at finite
string coupling this $SO(6)$ factor decomposes into a degeneration of
degree 4, and a degeneration of degree 1, separated by
$\Lambda_3$.

Let us try to understand the physics a bit better. In
order to do this it is convenient to use the classification of $(p,q)$
7-branes described in \cite{Gaberdiel:1997ud,DeWolfe:1998zf}. The
ordinary D7-branes are of type $(1,0)$, and are denoted as $A$-type
branes. An isolated $O7^-$ plane splits into two components in
F-theory, which can be denoted as $B$ and $C$, of $(p,q)$ type
$(-2,1)$ and $(0,1)$ respectively. The complete $SO(6)$ stack can then
be described as a $CBAAA$ stack. We would like to understand how the
five seven-branes in our stack split as we switch on a finite
coupling. In order to do this, it is convenient to consider the form
of the Seiberg-Witten curve for large (and equal) mass for the three
flavors. It is given by \cite{Seiberg:1994aj}:
\begin{align}
  y^2 = x^2(x-u) -\frac{1}{64}\Lambda_3^2(x-u)^2 -
  \frac{3}{64}m^2\Lambda_3^2(x-u) + \frac{1}{4}m^3\Lambda_3 x -
  \frac{3}{64}m^4\Lambda_3^2
\end{align}
As one would expect, this mass deformation separates the branes into
three stacks: three branes are located at $u_3=m^2+\Lambda_3m/8$, and
the two remaining branes are located at:
\begin{equation}
  u_{\pm} = \frac{1}{512}\left(\Lambda_3^2 - 96\Lambda_3 m \pm (\Lambda_3 + 64
    m)\sqrt{\Lambda_3^2+64\Lambda_3 m}\right)
\end{equation}
For large mass we can identify the branes at $u_\pm$ as the components
of the orientifold, and the stack of three branes as the three $A$
branes. Let us now smoothly take the mass to 0. For some intermediate
value of the mass the stack of three branes collides with the
degeneration at $u_+$, and the branes can have their $(p,q)$ labels
altered in the collision. After the collision the stack of three $A$
branes must become magnetic monopoles $(0,1)$ (so we recover a
fourplet of monopoles in the massless regime at $u=0$
\cite{Seiberg:1994aj}), while the brane at $u_+$ must become a
$(-1,2)$ dyon. This is indeed possible to achieve if we take the brane
at $u_+$ to be the $B$ brane, and we take the two brane stacks to
circle around each other once as they collide. We are left with $u_-$,
which was a spectator in the whole process, and which we identify as
the $C$ brane. We have depicted this process in
figure~\ref{fig:nf3monodromy}.

\FIGURE{
  \includegraphics[width=0.8\textwidth]{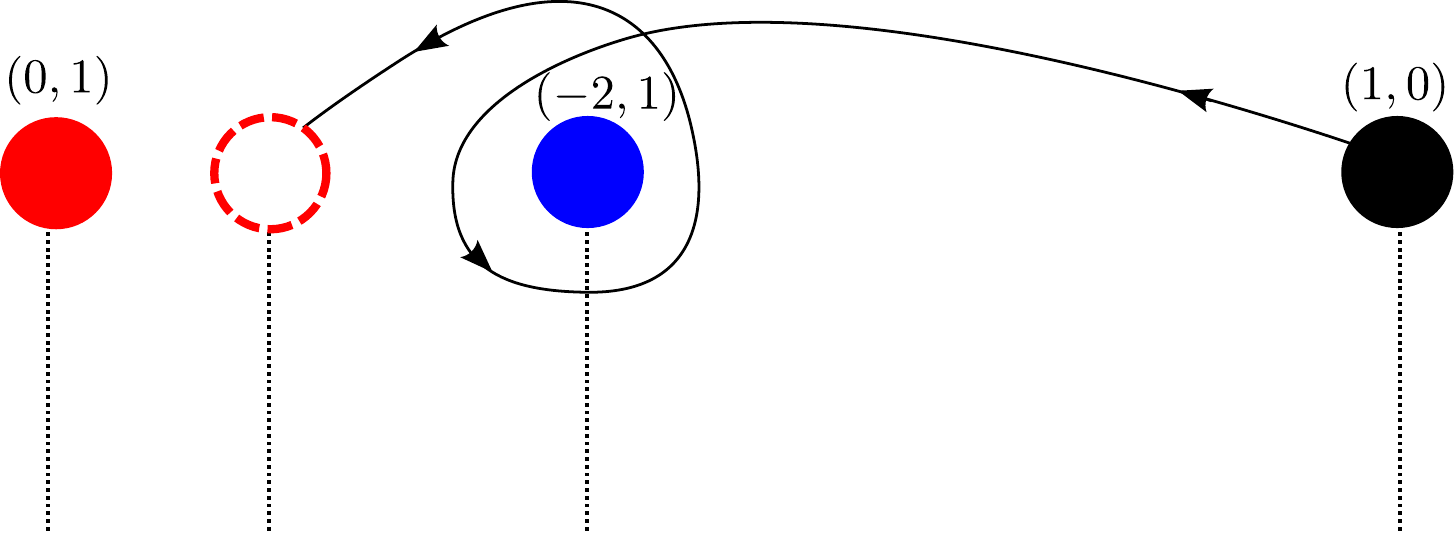}

  \label{fig:nf3monodromy}

  \caption{Schematic representation of the motion of the branes
    described in the text as we tune the mass parameter from large
    values down to 0. The red dot on the left represents the $C$ brane
    (of type $(0,1)$), the blue dot in the center the $B$ brane (type
    $(-2,1)$ before doing the monodromy), and the black dot on the
    right represents the stack of three $A$ branes (type $(1,0)$
    before monodromy). These $A$ branes become $(0,1)$ branes after
    the monodromy (shown as the dashed red dot), and the $B$ brane
    becomes a $(-1,2)$ dyon. We have indicated the branch cuts
    associated with the monodromies around each brane by the dotted
    line.}
}

In more detail, the process goes as follows: recall (from
\cite{Gaberdiel:1997ud}, for example) that a brane of type $(r,s)$
becomes a brane of type $(m,n)$ upon crossing the branch cut
associated with a $(p,q)$ brane, with:
\begin{equation}
  \begin{pmatrix}m\\n\end{pmatrix} =
  \begin{pmatrix}1-pq&p^2\\-q^2&1+pq\end{pmatrix}\begin{pmatrix}r\\s\end{pmatrix}.
\end{equation}
In figure~\ref{fig:nf3monodromy}, we have chosen conventions in in which
this is the monodromy for crossing the branch cut
counterclockwise. Denoting a brane of type $(p,q)$ as $X_{(p,q)}$, the
sequence of crossings in figure~\ref{fig:nf3monodromy} is then:
\begin{equation}
  CBA^3 \to C A^3 X_{(1,1)} \to C X_{(1,1)} X_{(0,1)}^3 \to C
  X_{(0,1)}^3 X_{(-1,2)} = C^4 X_{(-1,2)}
\end{equation}
So the local geometry is simply the one obtained from putting
four $C$ branes together, which gives an $SU(4)$ theory. The same result
can be obtained by studying the form of~\eqref{eq:SW-massless} close
to $u=0$.

\medskip

Let us briefly comment on what happens in various other interesting
configurations. If we tried to uplift $SO(4)$ stacks we would run into
the same phenomenon. The theory to study now is $\cN=2$ $SU(2)$ with 2
massless flavors. In this case the Seiberg-Witten curve is known to
degenerate at two points, both of degree 2
\cite{Seiberg:1994aj}. Using the same arguments as above, we can argue
that they correspond to a stack of two $(1,1)$ branes and a $CC$
stack. There is again a collision of stacks as we take the mass from 0
to large values, which changes the $(1,1)$ stack into an $AA$ stack,
and the $CC$ stack into a couple of neighboring $B$ and $C$
branes. Similarly, lifting a $SO(2)$ stack splits the configuration
into three separated degenerations of types $(1,1)$, $B$ and $C$.

One can argue in a similar fashion about what happens for most of the
other classical groups: $U(N)$ stacks induce $U(1)$ dynamics on the
probe, so no splitting occurs at finite coupling since the theory is
abelian, and thus IR-free. $Sp(N)$ stacks induce $SO(2)$ dynamics on
the probe, again non-confining. $SO(8)$ stacks give rise to a $\cN=2$
$N_f=4$ $SU(2)$ theory on the probe, which is conformal, so no IR
deformation of the geometry occurs. Similarly, $SO(2n)$ stacks with
$n>4$ give rise to IR-free theories on the probe.

\medskip

These results agree nicely with the Kodaira classification of
singularities, in that the classical stacks that split quantum
mechanically are exactly those that are missing from the
classification.\footnote{We reproduce the Kodaira classification in a
  form convenient for F-theory use in table~\ref{tab:TateTable}, in
  section~\ref{sec:tate}.} The analysis above also clarifies how some
of the string junctions found in \cite{DeWolfe:1998bi} for the BPS
states of $\cN=2$ $N_f<4$ can actually become massless in certain
points in moduli space, as Seiberg-Witten theory predicts they
should. As we discussed in detail above, the winding path connecting
the ``classical'' $A^{N_f}BC$ description of the flavor group and the
actual quantum configuration of D7 branes forces us to take the
effects of $SL(2,\bZ)$ monodromy into account. It is not hard to see
that the states in the classification of \cite{DeWolfe:1998bi} that
should become massless according to Seiberg-Witten theory, do indeed
``untangle'' due to the winding motion and the Hanany-Witten effect
\cite{Hanany:1996ie}, and go from being involved string junctions to
simple $(p,q)$-strings. These $(p,q)$ strings then become massless
when the D3 collides with the D7 branes.

\section{\v Cech Cohomology of Line Bundles over Toric Varieties}
\label{sec:cech}

In section~\ref{sec:zero-modes}, we will need to perform a calculation
of sheaf cohomology in order to show the absence of the
$\overline{\tau}_{\dot \alpha}$ mode for an $O(1)$ instanton. In this
section we explain in detail the steps involved in calculating such
cohomologies on toric varieties, and refer the reader to a code we
have written which performs such computations. For the sake of
brevity, we assume that the reader is familiar with the main concepts
used in the study of toric geometry, but highly recommend
\cite{Skarke:1998yk, Closset:2009sv,Bouchard:2007ik} for an
introduction and \cite{Cox:LecturesOnToricVarieties,
  CoxLittleSchenck:ToricVarieties, Fulton} for a thorough
treatment. For more details on \v Cech cohomology on toric varieties,
see Chapter 9 of \cite{CoxLittleSchenck:ToricVarieties}, which we
follow closely here.\footnote{The authors of
  \cite{CoxLittleSchenck:ToricVarieties} have kindly decided to
  provide recent copies of the book at the web address listed in the
  references, until it is completed and published by AMS.}

\subsection{General Discussion}
\label{sec:cech-general}

In general, the calculation of the \v Cech cohomology groups
$\check{H}(\cU,\cF)$ for a sheaf $\cF$ on $X$ requires knowledge of an
open cover $\cU$ of $X$, determination of the $p$th \v Cech cochains
$\check{C}^p(\cU,\cF)$, and determination of the differential maps
$d_p$, which are the maps between the \v Cech cochains in the \v Cech
complex
\begin{equation}
  \label{eq:cech-complex}
  0 \rightarrow \check{C}^0(\cU,\cF) \xrightarrow{d_0} \check{C}^1(\cU,\cF) \xrightarrow{d_1} \dots \xrightarrow{d_l} \check{C}^{l-1}(\cU,\cF) \xrightarrow{d_{l-1}} \check{C}^l(\cU,\cF) \rightarrow \ldots
\end{equation}
We will define the differentials in section~\ref{sec:dp1-cech} below.
The $p$th \v Cech cochains keep track of local sections, as can be
seen from the definition
\begin{align}
  \check{C}^p (\cU,\cF) & \equiv \bigoplus_{(i_0,\dots,i_p)\in[l]_p}
  \cF(U_{\sigma_{i_0}} \cap \dots \cap U_{\sigma_{i_p}}),\,\,\,\,\,
  l=|\cU |,
\end{align}
where $(i_0,\dots,i_p)\in[l]_p$ is a $(p+1)$-tuple of elements in the
set $[l]\equiv\{1,\dots,l\}$, which has the ordering
$i_0<\dots<i_p$. As $p$ increases, the sections become more and more
local, and the \v Cech complex can be viewed intuitively as encoding
how increasingly local sections ``fit together''. Given this data and
intuition, the $p$th \v Cech cohomology groups are defined to be
\begin{equation}
\check{H}^p(\cU,\cF) \equiv \frac{ker(d_p)}{im(d_{p-1})},
\end{equation}
as usual. After determining the structure of the $p$th \v Cech
cochains and the differential maps $d_p$, the \v Cech cohomology can
be computed directly as the cohomology of the
complex~\eqref{eq:cech-complex}.

In the generic case, however, the computation might be further
complicated by not knowing, a priori, an open cover of
$X$. Fortunately, in the case where $X$ is a toric variety $X_\Sigma$,
the affine toric variety $U_\sigma$ associated with a cone $\sigma$ is
a patch on $X_\Sigma$. Then there is a natural choice for an open
cover, namely
\begin{equation}
\cU \equiv \{U_\sigma\}_{\sigma \in \Sigma_{max}}, \,\,\,\,\, l=|\Sigma_{max}|,
\end{equation}
where $\Sigma_{max}$ is the set of top-dimensional cones. Moreover, to
determine the structure of the $p$th \v Cech cochain in general, we
must know the structure of $ \cF(U_{i_0}\cap\dots\cap U_{i_p})$, which
requires knowing how the opens in $\cU$ intersect. Again, it is a
fortunate property of toric varieties that the intersection of two
opens is encoded in the intersection of two cones. For example, if
$\sigma_1,\sigma_2\in \Sigma_{max}$ and $\tau = \sigma_1 \cap
\sigma_2$ is a common face, then
\begin{equation}
U_{\sigma_1} \cap U_{\sigma_2} = U_\tau.
\end{equation}
Thus, for toric varieties, the relevant intersections of opens are
known, and one can proceed directly to determining the structure of
the \v Cech cochains.

The cochains we are interested in are the \v Cech cochains of a sheaf
$\cO_{X_\Sigma}(D)$ on a toric variety $X_\Sigma$ with the natural
open cover $\cU$ on the toric variety. On an open patch $U_\sigma$
associated to some cone, not necessarily top-dimensional,
$\cO_{X_\Sigma}(D)(U)$ is an $\cO_{X_\Sigma}$-module finitely
generated by the set of monomials on $U_\sigma$ of class $[D]$ for the
divisor $D=\sum_\rho a_\rho D_\rho$. This just means that an arbitrary
$\alpha \in \cO_{X_\Sigma}(D)(U)$ is a linear combination of these
monomials with coefficients that are functions on $X_\Sigma$. The
monomials are local sections on the patch, so we write
\begin{align}
\label{eqn:Toric Cech Cochain}
\check{C}^p (\mathcal \cU,\mathcal O_{X_\Sigma}(D)) & = \bigoplus_{(i_0,\dots,i_p)\in[l]_p} H^0(U_{\sigma_{i_0}} \cap \dots \cap U_{\sigma_{i_p}},\mathcal O_{X_\Sigma}(D)).
\end{align}
Determining the local sections of class $[D]$ is not
difficult. Considering the fact that $\prod_\rho x_\rho^{a_\rho}$ has
class $[D]$ for $x_\rho$ the homogeneous coordinate associated with
the one-dimensional cone $\rho\in\Sigma(1)$, there is a monomial of
class $[D]$ for each $m\in M$, given by
\begin{equation}
\prod_\rho x_\rho^{\langle m,u_\rho\rangle + a_\rho},
\end{equation}
where $u_\rho\in N$ is the vector associated with the one-dimensional
cone $\rho$ and $\langle .,.\rangle$ is the dot product. It is of
class $[D]$ due to the fact that we have multiplied by a gauge
invariant product of homogeneous coordinates, $\prod_\rho x_\rho^{\langle m,u_\rho\rangle}$.

Calculating the structure of the \v Cech cochains involves determining
which of the monomials are well-defined on a given patch. For example,
if a monomial has $\langle m,u_\beta\rangle < -a_\beta$ for $\beta\in
\Sigma(1)$, then the monomial is only well defined on patches where
$x_\beta \ne 0$. This behavior is captured in a simple way by the
notion of ``+'' and ``-'' regions in the $M$ lattice, where the former
is  the halfplane $\langle m,u_\beta\rangle \ge -a_\beta$ and the latter is the halfplane $\langle
m,u_\beta\rangle < -a_\beta$. The $M$ lattice is then partitioned by
the set of lines $\langle m,u_\rho\rangle = -a_\rho \,\,\, \forall
\rho\in\Sigma(1)$, where each partition is a region in the $M$ lattice
categorized by a string of +'s and -'s, one for each homogeneous
coordinate. For example, on $\bP ^4$, a lattice point $m$ in the
region with sign ``$-++--$'' would have a corresponding monomial which is
only well-defined on patches where $x_1$, $x_4$, and $x_5$ are
non-zero. We will henceforth name such a region $R_{-++--}$, for the
sake of notation. How many lattice points are in this region, or
whether it exists at all, is highly dependent on the divisor $D$.

Given this intuition about local sections in terms of signed regions, we would like to relate them directly to patches $U_\sigma$, since we are interested in expressions of the form \eqref{eqn:Toric Cech Cochain}. We define
\begin{equation}
\label{eqn:Psigma}
P_\sigma=\{m\in M_\bR \,\,\,|\,\,\, \langle m,u_\rho\rangle \ge -a_\rho,\,\,\, \forall \rho\in\sigma(1)\},
\end{equation}
whose intersection with the $M$ lattice contains all lattice points $m$ whose corresponding monomials are local sections of $U_\sigma$. More precisely, 
\begin{equation}
\label{eqn:Local H0}
H^0(U_{\sigma_{i_0}} \cap \dots \cap U_{\sigma_{i_p}},\mathcal O_{X_\Sigma}(D)) = \bigoplus_{m\in P_{i_0\dots i_p}\cap M} \bC \cdot \chi^m,
\end{equation}
where $\chi^m$ and $P_{i_0\dots i_p}$ are shorthand for the monomial
corresponding to $m$ and $P_{\sigma_{i_0}\cap\dots\cap\sigma_{i_p}}$,
respectively. This identification makes sense in terms of patches,
because if $m\in P_\sigma\cap M$, then its corresponding monomial is
guaranteed to have positive exponent for the homogeneous coordinates
$x_\rho$ for all one-dimensional cones $\rho$ in $\sigma$. This is
necessary to be well-defined on $U_\sigma $, since
$D_\rho=\{x_\rho=0\}\subseteq U_\sigma$, $\forall \rho \in
\sigma(1)$. It is sufficient because $x_\rho\ne 0$ on $U_\sigma$ for
every $\rho\notin\sigma(1)$. One should note, of course, that a given
$P_\sigma$ is generically the union of multiple signed regions, and
moreover that a given signed region might contribute to multiple
$P_\sigma$ for different cones in the fan.

Having the requisite tools for explicitly constructing the \v Cech
cochains, it is straightforward to compute the
differentials\footnote{We do not give the general definition now,
  because we think it is more illustrative to state it when we
  will use it in the detailed $dP_1$ example.}, and one can then
directly compute the \v Cech cohomology groups
\begin{equation}
\check{H}^p(\cU,\cO_{X_\Sigma}(D)) \equiv \frac{ker(d_p)}{im(d_{p-1})}.
\end{equation}
The previous discussion was general but perhaps somewhat abstract. We
now proceed to illustrate how to apply these ideas in a simple but
non-trivial example, $dP_1$. As we will see, the \v Cech complex gives
a simple and systematic (albeit cumbersome, if done by hand) way to
compute line bundle cohomology.

\subsection{Calculating \v Cech Cohomology on $dP_1$}
\label{sec:dp1-cech}

As a concrete non-trivial example, we calculate an example of \v Cech
cohomology for a line bundle over the first del Pezzo surface,
$dP_1$. The del Pezzo surfaces are $\bP ^1\times \bP ^1$ and the
blow-up of $\bP ^2$ at $n$ points, $n=0,\dots,8$, which are denoted
$dP_n$. The fan which specifies $dP_1$ as a toric variety is given in
figure~\ref{fig:dP1fan}, and it is easy to see that the removal of
$u_4$, which corresponds to the exceptional divisor of the blow-up,
leaves us with the fan for $\bP ^2$. Hence, this is $dP_1$, also known
as the first Hirzebruch surface $\bF_1$.

\TABLE{
\centering
\hspace{-1cm}
\begin{minipage}{0.49\textwidth}
\centering
\scalebox{.6}{
\includegraphics[width=\textwidth]{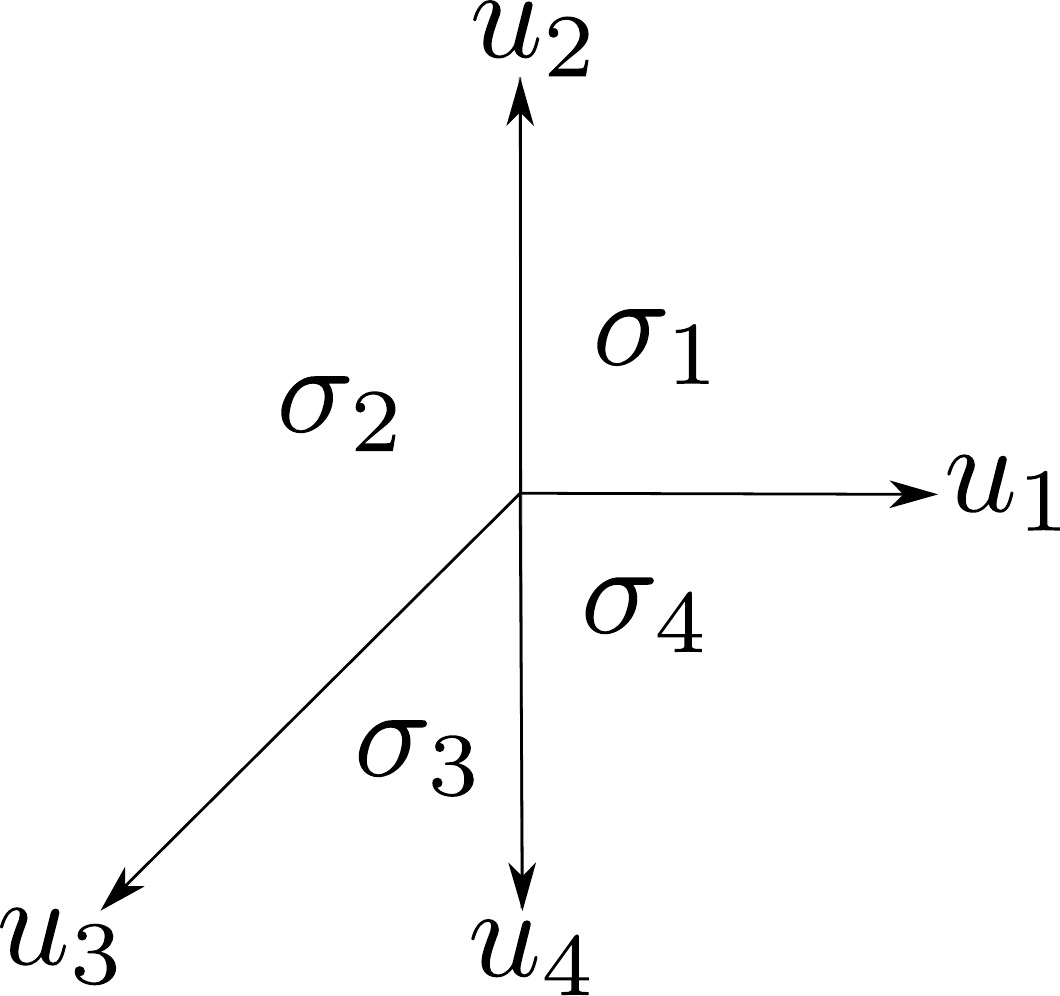}
}

\caption{The fan for $dP_1$. We have denoted the divisors $u_i$ and
  the top dimensional cones $\sigma_i$.}
\label{fig:dP1fan}

\end{minipage}
\hspace{0.6cm}
\begin{minipage}{0.49\textwidth}
\centering

 \begin{tabular}{r|r|cc|c} 
  Coords& Vertices & $Q^1$ & $Q^2$ & Divisor Class \\
  \hline
  $x$ & $u_1$=(1,0) & 1 & 0 & $H$ \\
  $y$ & $u_2$=(0,1) & 1 & 1 & $H+E$ \\
  $z$ & $u_3$=(-1,-1) & 1 & 0 & $H$ \\
  $w$ & $u_4$=(0,-1) & 0 & 1 & $E$ \\
  \hline 
  $\sum_i [D_i]$ & & 3 & 2 & $3H+2E$
  \end{tabular} 
  \caption{GLSM charges for $dP_1$.}
 \label{dP1GLSMTable} 

\end{minipage}
}

To fix notation, the homogeneous coordinates $x$, $y$, $z$, and $w$
are associated to the rays $u_1$, $u_2$, $u_3$, and $u_4$,
respectively. For this example, we choose to calculate the \v Cech
cohomology groups $\check{H}^p(\cU,\cO_{dP_1}(D))$ for the divisor
$D=5D_x-2D_w$. For this divisor, there are four lines which divide the
$M$ lattice into signed regions, given by
\begin{align}
  \begin{split}
    l_1:\,\,\,m_x=-5,\qquad 
    &l_2:\,\,\,m_y=0 \\
    l_3:\,\,\,m_x+m_y=0,\qquad
    &l_4:\,\,\,m_y=-2,
  \end{split}
\end{align}
which correspond to the rays $u_1$, $u_2$, $u_3$, and $u_4$,
respectively. The partitioned $M$ lattice is given in
figure~\ref{fig:dP1regions}, where each region has been labeled with
the appropriate sign according to the conventions discussed in the
previous section.

\FIGURE{
\begin{minipage}{0.6\textwidth}
\centering
\includegraphics[width=0.6\textwidth]{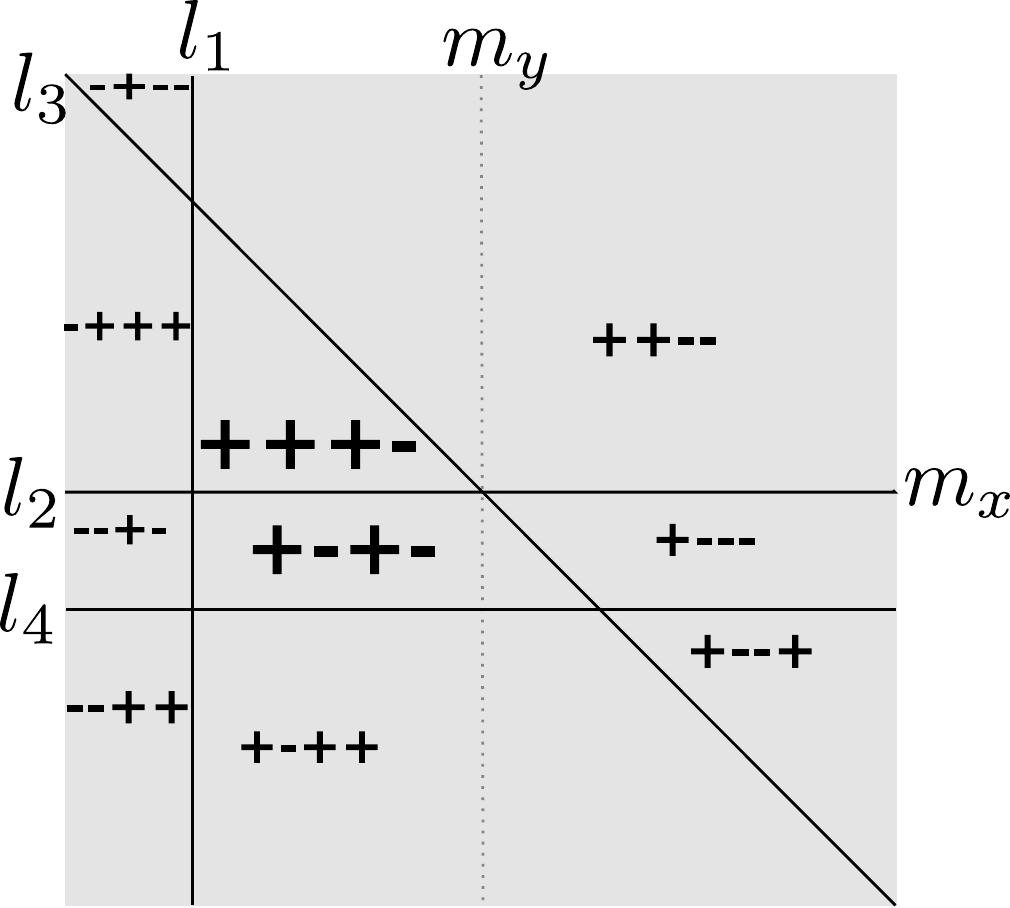}
\end{minipage}

\caption{Signed regions in the $M$ lattice corresponding to $\mathcal
  O(5D_x-2D_w)$ over $dP_1$. We have denoted by $m_x$ and $m_y$ the
  coordinate axes of the $M$ lattice.}
\label{fig:dP1regions}
}

Calculationally, rather than considering which signed regions have
monomials well-defined on the intersection of a particular set of
opens, it is useful to instead consider on which intersections of
opens a particular monomial is well-defined. In the end, this
essentially corresponds to considering the cohomological contribution
of each point in the $M$ lattice. All $m$ in a given signed region
will have the same contribution. This is useful since each point in
the $M$ lattice contributes independently to the cohomology. In other
words, there is a grading on cohomology which allows us to consider
the contribution of each $m\in M$ independently. We refer the reader to chapter
9 of \cite{CoxLittleSchenck:ToricVarieties} for more details.

For this reason, we would like to categorize those $P_\sigma$'s which
contain the $m$'s corresponding to monomials well defined on a
particular intersection, as a union of signed regions. The result is
\begin{align}
\begin{split}
\label{eqn:Psigmas}
P_1 = \bigcup R_{++\bullet\bullet} \,\,\,\,\,\,\,\,\,\, P_2 = \bigcup
R_{\bullet ++\bullet} \,\,\,\,\,\,\,\,\,\,
P_3 = \bigcup R_{\bullet\bullet ++} \,\,\,\,\,\,\,\,\,\, P_4 = \bigcup R_{+\bullet\bullet +} \\
P_{12} = \bigcup R_{\bullet +\bullet\bullet} \,\,\,\,\,\,\,\,\,\, P_{13} = \bigcup R_{\bullet\bullet\bullet\bullet} \,\,\,\,\,\,\,\,\,\, P_{14} = \bigcup R_{+\bullet\bullet\bullet}   \\
P_{23} = \bigcup R_{\bullet\bullet +\bullet} \,\,\,\,\,\,\,\,\,\, P_{24} = \bigcup R_{\bullet\bullet\bullet\bullet} \,\,\,\,\,\,\,\,\,\, P_{34} = \bigcup R_{\bullet\bullet\bullet +}  \\ 
P_{123} = \bigcup R_{\bullet\bullet\bullet\bullet}
\,\,\,\,\,\,\,\,\,\, P_{124} = \bigcup
R_{\bullet\bullet\bullet\bullet} \,\,\,\,\,\,\,\,\,\, P_{134} =
\bigcup R_{\bullet\bullet\bullet\bullet} \,\,\,\,\,\,\,\,\,\,
P_{234} = \bigcup R_{\bullet\bullet\bullet\bullet}  \\
P_{1234} = \bigcup R_{\bullet\bullet\bullet\bullet}, 
\end{split}
\end{align}
where a $\bullet $ simply means that the union includes both the $+$
and the $-$ in that placeholder, so that $\cup R_{+\bullet\bullet} =
R_{+++} \cup R_{++-} \cup R_{+-+} \cup R_{+--}$. This allows us to
consider the contributions of a particular $m\in M$ to a \v Cech
cochain as a vector where different entries correspond to different
intersections of opens. Examples will come when we do the actual
calculation.

The only technical aspect which must still be specified before
actually computing the kernels and images of the differentials $d_p$
is the definition and form of the differentials themselves. In
general, they are maps from $\check{C}^p(\cU,\cF)$ to
$\check{C}^{p+1}(\cU,\cF)$ defined by
\begin{equation}
  \label{eq:cech-differential}
(d_p\sigma)_{i_0\dots i_{p+1}} = \sum_{k=0}^{p+1} (-1)^k \sigma_{i_0\dots \hat{i}_l \dots i_{p+1}}|_{U_{i_0}\cap\dots\cap U_{i_{p+1}}},
\end{equation}
where $\hat{i}_k$ indicates that this index is removed. For a given
set of indices $(i_0,\dots,i_{p+1})$, this specifies one
  component in an element of $\check{C}^{p+1}(\cU,\cF)$. As an example, the
definition~\eqref{eq:cech-differential} gives
\begin{equation}
\label{eqn:dpexample}
(d_1\sigma)_{134} = \sigma_{34}|_{U_1\cap U_3 \cap U_4} - \sigma_{14}|_{U_1\cap U_3 \cap U_4} + \sigma_{13}|_{U_1\cap U_3 \cap U_4}
\end{equation}
for the case where $l=4$, which is our case for $dP_1$. Each component
in an element of a \v Cech cochain is specified by a $(p+1)$-tuple of indices, where
the components are ordered in a vector according to the natural
ordering on $[l]_p$. Thus, equation \eqref{eqn:dpexample} corresponds
precisely to the third row in $d_1$, listed below.

All components of $d_p\sigma$ can be determined this way, which allows
us to write the maps as matrices. The result in our particular case is
\begin{align}
  0 \rightarrow \check{C}^0(\cU,\cO_{dP_1}(5D_x-2D_w))
  \xrightarrow{d_0= \left( \scalebox{.6}{
\begin{tabular}{cccc}
-1 & 1 & 0 & 0\\
-1 & 0 & 1 & 0\\
0 & -1 & 1 & 0\\
0 & -1 & 1 & 0\\
0 & -1 & 0 & 1\\
0 & 0 & -1 & 0
\end{tabular}} \right)
} & \check{C}^1(\cU,\cO_{dP_1}(5D_x-2D_w)) \\ \notag  \xrightarrow{d_1=	\left( \scalebox{.6}{
\begin{tabular}{ccccccc}
1 & -1 & 0 & 1 & 0 & 0\\
1 & 0 & -1 & 0 & 1 & 0\\
0 & 1 & -1 & 0 & 0 & 1\\
0 & 0 & 0& 1 & -1 & 1
\end{tabular}} \right)}  &\check{C}^2(\cU,\cO_{dP_1}(5D_x-2D_w)) \\ \notag  \xrightarrow{d_2=	\left( \scalebox{.6}{
\begin{tabular}{cccc}
-1,1,-1,1
\end{tabular}} \right)}&
\check{C}^3(\cU,\cO_{dP_1}(5D_x-2D_w)) \rightarrow \ldots
\end{align}
Notice that the definition of the differential, seen as a linear map
between vector spaces, does not require us to specify which monomial we
are dealing with. This information will only enter in the definition
of the vector spaces $\check{C}^\bullet(\cU,\cO_{dP_1}(5D_x-2D_w))$,
specifying which elements of the vector space are necessarily
vanishing due to the monomial under consideration not being well
defined in the relevant patch.

Now all of the pieces are in place for a direct computation of
cohomology. We emphasize again that it is sufficient to consider the
cohomology corresponding to a given $m\in M$, and then sum over the
contributions from each $m$. Moreover, since all $m$'s in a given
signed region contribute to the overall cohomology in the same way, it
is only necessary to compute the cohomological contributions for each
signed region and to then multiply that contribution by the number of
points in that region. This implies that all cohomological
contributions from non-compact regions must be zero, since there are
an infinite number of points, and the cohomology is finite. This means
that we only need to calculate the contributions from points in
$R_{+++-}$ and $R_{+-+-}$.

Let us study first the monomials in $R_{+-+-}$. From equation
\eqref{eqn:Psigmas} and the natural ordering of $(p+1)$-tuples in
$[l]_p$, elements of the \v Cech cochains for a given $m$ can be
written
\begin{align}
\begin{pmatrix}
0 \\
0 \\
0 \\
0 
\end{pmatrix}
\cdot \chi^m \in \check{C}^0(\cU,\cO_{dP_1}(5D_x-2D_w)), \,\,\,\,\, 
\begin{pmatrix}
0 \\
a \\
b \\
c \\
d \\
0
\end{pmatrix}
\cdot \chi^m \in \check{C}^1(\cU,\cO_{dP_1}(5D_x-2D_w)), \,\,\,\,\, \\ \notag
\begin{pmatrix}
e \\
f \\
g \\
h 
\end{pmatrix}
\cdot \chi^m \in \check{C}^2(\cU,\cO_{dP_1}(5D_x-2D_w)), \,\,\,\,\, 
\begin{pmatrix}
i
\end{pmatrix}
\cdot \chi^m \in \check{C}^3(\cU,\cO_{dP_1}(5D_x-2D_w)), \,\,\,\,\,
\end{align}
where $a,b,c,d,e,f,g,h,i\in \bC$. One can then consider the action of the appropriate $d_p$'s on the
these elements, and it is a straightforward exercise in linear algebra
to show that all of the kernels and images are the same except for
$im(d_0)=0$, $ker(d_1)=\bC$. Thus, for each $m$ in this region, the
contribution is $\check{h}^\bullet_m = (0,1,0)$.

\FIGURE{
\includegraphics[width=0.3\textwidth]{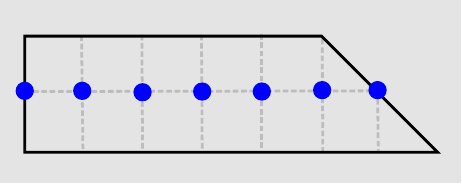}
\caption{The only signed region which contributes to the cohomology.}
\label{fig:dP1relregion}
}
In order to count points in this region recall from the definition
of signed regions that ``+'''s are inclusive while ``-'''s are
exclusive. With this in mind, only the filled dots in
figure~\ref{fig:dP1relregion} contribute. Thus, the contribution
of this region to the cohomology is given by
\begin{equation}
\check{h}^\bullet_{R_{+-+-}}(\cU,\cO_{dP_1}(5D_x-2D_w))=(0,7,0).
\end{equation}

A similar argument in the $R_{+++-}$ region shows that it does not
contribute to the cohomology, and thus we conclude that
\begin{equation}
\check{h}^\bullet(\cU,\cO_{dP_1}(5D_x-2D_w))=(0,7,0).
\end{equation}

\subsection[The Koszul Complex]{The Koszul Complex\footnote{We would
    like to acknowledge a number of useful discussions with
    L. Anderson on the contents of this section.}}
\label{sec:koszul}

Once we know the cohomology of line bundles on the ambient space, we
can use an exact sequence known as the Koszul complex to obtain the
cohomology on subspaces of this ambient space. Let us start with the
case of induced line bundles on divisors of $\cA$. Denoting as $N^*$
the dual of the normal bundle of our surface $X$ on $\cA$, we have
that:
\begin{equation}
  0\to N^* \to \cO_\cA \to \cO_X \to 0 
\end{equation}
This formula does not require that $\cA$ is toric. In the case of a divisor $D$ of $\cA$, $N^*$ is the line bundle $\cO(-D)$. Furthermore, in order
to obtain information about the cohomology of a line bundle $\cL$ on
$X$, we can tensor the whole short exact sequence above by $\cL$, we
get:
\begin{equation}
  0\to \cO(\cL-D) \to \cO_\cA(\cL) \to \cO_X(\cL) \to 0   
\end{equation}
Any short exact sequence gives rise to a long exact sequence in
cohomology in a standard way. In our particular case we get:
\begin{align}
  \begin{split}
  0 & \to H^0(\cA, \cO(\cL - D)) \to H^0(\cA, \cO(\cL)) \to H^0(X,
  \cO(\cL)) \to \\
  &\to H^1(\cA, \cO(\cL - D)) \to H^1(\cA, \cO(\cL)) \to H^1(X,
  \cO(\cL)) \to \\
  &\to \ldots \to H^d(\cA, \cO(\cL - D)) \to H^d(\cA, \cO(\cL)) \to H^d(X,
  \cO(\cL)) \to 0
  \end{split}
\end{align}
where $d$ is the dimension of the ambient space.

From here we can read the dimensions of the cohomology groups. A
couple of very useful facts are that we can always split any exact
sequence
\begin{equation}
  0\to A \to B \to C \to D \to \ldots
\end{equation}
into two pieces:
\begin{align}
  0 \to A \to B \to X \to 0\\
  0 \to X \to C \to D \to \ldots
\end{align}
and that for any short exact sequence $0\to A \to B \to C\to 0$ we
have
\begin{equation}
  \dim(B) = \dim(A)+\dim(C),
\end{equation}
which allows one to compute the dimensions of the cohomologies in a straightforward manner.

For the case of a complete intersection of three divisors in the
ambient space (our case in the main text), there is a useful general
form for the Koszul complex, given by:
\begin{equation}
0 \rightarrow \wedge^3N^* \rightarrow \wedge^2N^* \rightarrow N^* \rightarrow \cO_\cA \rightarrow \cO_{D_1\cap D_2\cap D_3} \rightarrow 0,
\end{equation}
where $N$ is the sum of the normal bundles of the divisors, $N\equiv
N_{D_1}\oplus N_{D_2} \oplus N_{D_3}$. After splitting this sequence
into short exact sequences, one uses those sequences to arrive at a
number of long exact sequences in cohomology, which make it
straightforward to compute the relevant groups. For the complete
intersection of more hypersurfaces, the above sequence extends as one
might expect.

\subsection{Computer Implementation}
\label{sec:computer implementation}

We have implemented the algorithm described in
section~\ref{sec:cech-general} using a combination of SAGE
\cite{sage}, C code and code from the Computational Homology Project
\cite{chomp}.

The code and accompanying documentation can be downloaded at the web
address:
\begin{center}
\href{http://www.sas.upenn.edu/~inaki/Cech.html}{http://www.sas.upenn.edu/~inaki/cech.html}
\end{center}

\section{Example: a $10\,10\,5_H$ Yukawa Coupling in F-theory}
\label{sec:example}

In this section we will illustrate the previous considerations in a
particular example of phenomenological interest. The discussion is
organized as follows: in section \ref{sec:IIB}, we present the
geometric data for a IIB GUT orientifold compactification on a
Calabi-Yau manifold realized as a hypersurface in a toric
variety. This IIB background has an euclidean D3-brane instanton
generating a $10\,10\,5_H$ coupling. In section \ref{sec:fourfold}, we
lift this IIB model to F-theory by specifying an elliptically fibered
Calabi-Yau fourfold as a complete intersection in a six-dimensional
toric variety. In section \ref{sec:zero-modes}, we study the lifted
(now M5) instanton generating the $10\,10\,5_H$ coupling. By an
explicit line bundle cohomology computation we show the absence of
fermionic zero modes that would make the contribution of the instanton
to the superpotential vanish. In section \ref{sec:tate}, we present
the precise form of the Tate sections and show that many features of
the gauge $D7$ branes and $O7$ planes in IIB can be seen in the lift,
but in agreement with the discussion in
section~\ref{sec:quantum-splitting} the $SO(6)$ gauge stack cannot be
obtained at nonzero $g_s$.

\subsection{The IIB Geometry}
\label{sec:IIB}
For the sake of reference, in this section we present the geometric
data relevant for the F-theory lift of the manifold $M_2^{({\rm
    dP}_9)^2}$, henceforth called $X$, which is a Calabi-Yau threefold hypersurface in a
four-dimensional toric variety. It was presented in
\cite{Blumenhagen:2008zz} as a suitable manifold for $SU(5)$ GUT model
building in IIB orientifold compactifications. It exhibits many
desirable features, including the generation of the $10\,10\,5_H$ Yukawa
coupling via a euclidean $D3$ instanton.

The GLSM charges representing the ambient toric variety $X_\Sigma$ are
given in table~\ref{IIBGLSMTable}. As required by the Calabi-Yau
condition, the hypersurface X has divisor class
equal to the anticanonical class of the ambient toric variety, that is
$\sum_i \, [D_i]$.
\TABLE{
\centering
 \begin{tabular}{r|cccc|c} 
  Coords / Vertices& $Q^1$ & $Q^2$ & $Q^3$ & $Q^4$ & Divisor Class \\
  \hline
  $x_1=(1, 0, 0, 0)$ & 3 & 0 & 0 & 0 & $3M$ \\
  $x_2=(0, 1, 0, 0)$ & 2 & 0 & 0 & 0 & $2M$ \\
  $x_3=(0, 0, 1, 0)$ & 0 & 1 & 0 & 0 & $N$ \\
  $x_4=(0, 0, 0, 1)$ & 0 & 0 & 1 & 0 & $O$ \\
  $x_5=(-9, -6, -1, -1)$ & 0 & 1 & 1 & -1 & $N\,+\,O\,-\,P$ \\
  $x_6=(-3, -2, 0, 0)$ & 1 & -1 & -1 & -1 & $M\,-\,N\,-\,O\,-\,P$ \\
  $x_7=(-6, -4, -1, 0)$ & 0 & 0 & -1 & 1 & $-O\,+\,P$ \\
  $x_8=(-6, -4, 0, -1)$ & 0 & -1 & 0 & 1 & $-N\,+\,P$ \\
  \hline 
  $\sum_i [D_i]$ & 6 & 0 & 0 & 0 & $6M$
  \end{tabular} 
  \caption{GLSM charges for $X_\Sigma$, the ambient toric variety on
    the IIB side whose Calabi-Yau hypersurface is the threefold X. We
    have chosen the basis of linearly inequivalent divisors
    $(M,N,O,P)$ to have charges
    $(1,0,0,0),(0,1,0,0),(0,0,1,0),(0,0,0,1)$ under the indicated
    $\bC^*$ gauge symmetries $Q^i$ of the GLSM. We have also
    indicated next to each field $x_i$ the corresponding
    one-dimensional generator of the fan.}
 \label{IIBGLSMTable} 
}
In addition to this information, the orientifold involution $\sigma$
is taken to be
\begin{equation}
\sigma:\,\,\,x_3 \mapsto -x_3,
\end{equation}
under which the divisors $D_3$ and $D_7$ are fixed.\footnote{$D_i$ is
  the vanishing locus of the homogeneous coordinate $x_i$, which can
  be written in terms of the generators of the divisor group, as in
  table~\ref{IIBGLSMTable}.} This identifies them as $O7$-planes, so
that $[O7]=[D_3]+[D_7]$. Furthermore, via projective
equivalences it can be seen that the points $x_5=x_6=x_8=0$ and
$x_1=x_5=x_8=0$ are fixed points of the $\sigma$-action, and thus are
the locations of $O3$-planes.\footnote{As noted in
  \cite{Blumenhagen:2008zz}, the points $x_4=x_7=x_8=0$ and
  $x_4=x_5=x_6=0$ are also fixed under the involution. However, the
  monomials $x_4x_5$ and $x_4x_8$ are in the Stanley-Reisner ideal,
  and thus these points are not in $X_\Sigma$. This is equivalent to
  them being in the set $Z_\Sigma$ in the homogeneous coordinate
  construction of this toric variety,
  $X_\Sigma=(\field{C}^8-Z_\Sigma)/G.$}

The intersection ring on the base can be computed using standard
techniques of algebraic geometry. In order to talk sensibly about
intersections we need to give a triangulation of the fan, or
equivalently the Stanley-Reisner ideal (loosely speaking, the set of
monomials in which not all terms can vanish simultaneously). We choose
the following simplicial triangulation:
\begin{align}
  \begin{split}
    [[1, 2, 3, 4], [1, 2, 3, 8], [1, 2, 4, 7], [1, 2, 5, 7],
    [1, 2, 5, 8], [1, 3, 4, 6],\\ [1, 3, 6, 8], [1, 4, 6, 7],
    [1, 5, 6, 7], [1, 5, 6, 8],
    [2, 3, 4, 6], [2, 3, 6, 8],\\
    [2, 4, 6, 7], [2, 5, 6, 7], [2, 5, 6, 8]]
  \end{split}
\end{align}
where the integers refer to generators of the fan (so $n$ stands for
$x_n$). The corresponding Stanley-Reisner ideal is
\begin{equation}
  SRI = \{x_3x_5, x_3x_7, x_4x_5, x_4x_8, x_7x_8, x_1x_2x_6\},
\end{equation}
which can be seen directly from the triangulation. For example, since there is no cone
in the triangulation with both $3$ and $5$, we know that $x_3x_5$ is in the Stanley-Reisner ideal.

In fact, this variety
was analyzed in \cite{Blumenhagen:2008zz} using a different base of
divisors. We can just take the result quoted there and change the
base to our $M,N,O,P$ cycles. The change of base is the following:
\begin{align}
  \begin{split}
    M & = \frac{1}{2}D_2 = 3D_5 + D_6 + 2D_7 + 2D_8\\
    N & = D_3 = D_5 + D_7\\
    O & = D_4 = D_5 + D_8\\
    P & = N+O-D_5 = D_5 + D_7 + D_8
  \end{split}
\end{align}
Here we have defined the basic divisors $D_i$ as the ones given by
$x_i=0$, and we have used the linear equivalences of
divisors:\footnote{We also list the linear equivalence relation for
  $D_1$, although it is not necessary for our calculations above.}
\begin{align}
  \begin{split}
    D_1 & = 9D_5 + 3D_6 + 6D_7 + 6 D_8\\
    D_2 & = 6D_5 + 2D_6 + 4D_7 + 4 D_8\\
    D_3 & = D_5 + D_7\\
    D_4 & = D_5 + D_8
  \end{split}
\end{align}

Using this change of basis, and the intersection form given in
\cite{Blumenhagen:2008zz}:
\begin{equation}
  I_X = D_6(7D_6^2 - D_5^2 - D_7^2 - D_8^2 - D_5D_6 - D_6D_7 - D_6D_8
  + D_5D_7 + D_5 D_8)
\end{equation}
we can easily obtain the triple intersection form in terms of
$M,N,O,P$. The result is:
\begin{equation}
  I_X = M(7M^2 + 2MN + 2MO + 3MP + NO + NP + OP + PP)
\end{equation}

\subsection{The Uplift to F-theory}
\label{sec:fourfold}

In this section we present the lift of the IIB orientifold model to
F-theory, where the geometry is that of an elliptically fibered
Calabi-Yau fourfold $Y$ of the form
\begin{equation}
T^2\,\longhookrightarrow Y \, \longtwoheadedrightarrow X/\sigma.
\end{equation}
We follow the prescription of
\cite{Collinucci:2008zs,Collinucci:2009uh,Blumenhagen:2009up}, which
is generalizable to many lifts of IIB orientifolds, and discuss the
details of this particular lift.\footnote{There has also been great
  progress recently in constructing semi-realistic global models
  directly in F-theory
  \cite{Marsano:2009ym,Marsano:2009gv,Blumenhagen:2009yv,Marsano:2009wr,Grimm:2009yu}.}
We construct first the base of the elliptic fibration $X/\sigma$ as a
hypersurface in a new ambient toric variety $X_{\Sigma^{'}}$, with
homogeneous coordinates whose GLSM charges have been changed relative
to their counterparts in $X_\Sigma$ to account for modding out by the
orientifold action. We then determine the divisor class of the
hypersurface $X/\sigma$ and use it to calculate the canonical bundle
of the base, $K_{X/\sigma}$, which is crucial in determining the
precise form of the Tate sections $a_n$. Next, having relevant
knowledge of the fourfold base, we construct a six-dimensional toric
variety $X_{\Sigma^{''}}$, in which the Calabi-Yau fourfold $Y$ is a
complete intersection of two hypersurfaces, one for the base and one
for the fiber. The GLSM charges for the homogeneous coordinates of the
base carry over from the toric variety $X_{\Sigma^{'}}$, and we show
how to determine the GLSM charges for the fiber-related coordinates
$x$, $y$, and $z$ from the Weierstrass equation. We also briefly
mention how one could arrive at the ambient toric variety of the
fourfold $X_{\Sigma^{''}}$ without explicitly constructing the
intermediate toric variety $X_{\Sigma^{'}}$.

\subsubsection*{The Base of $Y$}

Since we would like to stay in the framework of toric geometry, we
will start by constructing a toric ambient space for the
base. Specifically, the Calabi-Yau threefold $X$ on the IIB side is a
hypersurface in the toric variety $X_\Sigma$, so that one can
construct the base of the fourfold by modding out by the orientifold
action, giving a new toric ambient space $X_{\Sigma^{'}}$, and by
mapping the hypersurface constraints appropriately. This requires a
map from $X_\Sigma$ to $X_{\Sigma^{'}}$ which is $2$-to-$1$ away from
the $O7$-planes and $1$-to-$1$ on them. We choose the map to be
\begin{align}
\label{eqn:coordinatemap}
(x_1,x_2,x_3,x_4,x_5,x_6,x_7,x_8) \mapsto
(x_1,x_2,x_3^2,x_4,x_5,x_6,x_7^2,x_8) \\ \equiv (\tilde x_1, \tilde
x_2, \tilde x_3, \tilde x_4, \tilde x_5, \tilde x_6, \tilde x_7,
\tilde x_8),\notag
\end{align}
where the latter are the homogeneous coordinates of
$X_{\Sigma^{'}}$. The effect of such a map is a simple doubling of the
GLSM charges of $\tilde x_3$ and $\tilde x_7$ relative to $x_3$ and
$x_7$, while the charges of the other $\tilde x_i$ are left
unchanged. This is sufficient to determine the toric data of
$X_{\Sigma^{'}}$ presented in table~\ref{FBaseGLSMTable}.

\TABLE{
 \begin{tabular}{c|cccc|c} 
  Coords/Vertices& $Q^1$ & $Q^2$ & $Q^3$ & $Q^4$ & Divisor Class \\
  \hline
  $\tilde x_1=(1, 0, 0, 0)$ & 3 & 0 & 0 & 0 & $3I$ \\
  $\tilde x_2=(0, 1, 0, 0)$ & 2 & 0 & 0 & 0 & $2I$ \\
  $\tilde x_3=(0, 0, 1, 0)$ & 0 & 2 & 0 & 0 & $2J$ \\
  $\tilde x_4=(0, 0, 0, 1)$ & 0 & 0 & 1 & 0 & $K$ \\
  $\tilde x_5=(-9, -6, -2, -1)$ & 0 & 1 & 1 & -1 & $J\,+\,K\,-\,L$ \\
  $\tilde x_6=(-3, -2, 0, 0)$ & 1 & -1 & -1 & -1 & $I\,-\,J\,-\,K\,-\,L$ \\
  $\tilde x_7=(-3, -2, -1, 0)$ & 0 & 0 & -2 & 2 & $-2K\,+\,2L$ \\
  $\tilde x_8=(-6, -4, 0, -1)$ & 0 & -1 & 0 & 1 & $-J\,+\,L$ \\
  \hline 
  $\sum_i [\tilde D_i]$ & 6 & 1 & -1 & 1 & $6I\,+\,J\,-\,K\,+\,L$
  \end{tabular} 
  \label{FBaseGLSMTable}
  \caption{GLSM Charges for $X_{\Sigma^{'}}$, the four-dimensional ambient toric
    variety for the base $X/\sigma$ of the elliptic fibration on the
    F-theory side. We have indicated the generators of the fan.}
}

Having deduced the GLSM charges for the homogeneous coordinates $\tilde x_i$ in $X_{\Sigma^{'}}$, we must also deduce the divisor
class of $X/\sigma$. To this end, the divisor class of $X$ in
$X_\Sigma$ is $\sum_i [D_i] = 6I$. Monomials of this divisor class in
$X_\Sigma$ get mapped to monomials of base coordinates in $X_{\Sigma^{'}}$
via the map \eqref{eqn:coordinatemap}, from which we can read off the
divisor class of $X/\sigma$ in $X_{\Sigma^{'}}$. For example, from
\begin{equation}
x_5^{18}\,x_6^{6}\,x_7^{12}\,x_8^{12} \,\mapsto\, \tilde x_5^{18}\,\tilde x_6^{6}\,\tilde x_7^{6}\,\tilde x_8^{12}
\end{equation}
we see that $X/\sigma$ has class $6I$. From this, the anticanonical
bundle of the base can be computed from the adjunction formula to be
$\overline{K}_{X/\sigma}=c_1(T_{X/\sigma})=\sum_i \tilde D_i - 6I = J
- K + L$. Thus, we see that $X/\sigma$ is not Calabi-Yau.

At this point, one could explicitly construct the Tate form of the
elliptic fibration, since it is specified by sections $a_n\in
H^0(X/\sigma,K_{X/\sigma}^{-n})$, and we have calculated the divisor
class of the anticanonical bundle. This method was employed in
\cite{Blumenhagen:2009up} and was fruitful in examining the gauge
enhancements associated with fiber degenerations. However, since we
are interested in counting instanton zero modes via cohomologies of a
divisor wrapped by a vertical $M5$ brane, it is useful to construct
the full elliptically fibered fourfold $Y$ as a complete intersection
in a toric ambient space. In doing so, we will be able to apply the
algorithm descibed in section~\ref{sec:cech} in a straightforward
manner.

\subsubsection*{The Elliptically-Fibered Fourfold Y}

The process of constructing the ambient toric variety
$X_{\Sigma^{''}}$ of the fourfold $Y$ is fairly intuitive, as one might
expect, and essentially amounts to appropriately adding homogeneous
coordinates for the fiber. In addition, since we wish to realize the
fourfold as a complete intersection
\begin{equation}
Y\equiv\{P_{X/\sigma}=0\}\cap\{P_{T^2}=0\},
\end{equation}
we must specify the divisor class of the polynomials $P_{X/\sigma}$
and $P_{T^2}$. The polynomial $P_{T^2}$ is usually chosen to be in
either the Weierstrass form or (equivalently) the Tate form for an
elliptic curve. For ease in determining the relevant GLSM data, we
will use the Weierstrass form in this section, but will later move to
the Tate form to make the determination of gauge enhancements more
tractable. There is no technical difference, of course, since the Tate
sections determine $f$ and $g$. We merely choose one or the other
based on what is easiest for the particular task at hand.

Beginning with the base, the GLSM charges for the homogeneous
coordinates in $X_{\Sigma^{'}}$ carry over directly to
$X_{\Sigma^{''}}$, with the addition of the fact that they are
uncharged under the GLSM charge associated with the fiber, $Q^5$. We
immediately know that $[P_{X/\sigma}]=6I$, since it is must have the
same divisor class as $X/\sigma$ in $X_{\Sigma^{'}}$.

In addition to the polynomial $P_{X/\sigma}$, we must take into
account a polynomial $P_{T^2}$ corresponding to the elliptic fiber. As
mentioned above, in this section we choose the form
\begin{equation}
P_{T^2}\equiv y^2-x^3-fxz^4-gz^6,
\end{equation}
the vanishing locus of which gives an elliptic curve in Weierstrass
form, where $f$ and $g$ are global sections $f\in
H^0(X/\sigma,K_{X/\sigma}^{-4})$ and $g\in
H^0(X/\sigma,K_{X/\sigma}^{-6})$. In the case where $f$ and $g$ are
merely complex numbers, rather than sections, the Weierstrass equation
can be considered to be a degree six hypersurface in
$\bP_{2,3,1}$. This gives the charges $Q^5$ of $x$, $y$, and $z$ under
the projective scaling associated only to the fiber
coordinates. Moreover, from homogeneity of the Weierstrass equation,
the classes $[D_x]$ and $[D_y]$ can be determined as
\begin{equation}
2[D_y]=3[D_x]=[g]+6[D_z],
\end{equation}
where we use $[g]=6[\overline{K}_{X/\sigma}]=6J-6K+6L$. In addition,
since we have two equations and three unknowns, we choose $[D_z]=M$, so
that it does not transform under projective scalings of the base. This
is sufficient to determine the toric data of $X_{\Sigma^{''}}$ presented
in table~\ref{FGLSMTable}.

\TABLE{\centering
 \begin{tabular}{c|ccccc|c} 
  Coords / Vertices& $Q^1$ & $Q^2$ & $Q^3$ & $Q^4$ & $Q^5$ & Divisor Class \\
  \hline
  $\tilde x_1=(1, 0, 0, 0, 0, 0)$ & 3 & 0 & 0 & 0 & 0 & $3I$ \\
  $\tilde x_2=(0, 1, 0, 0, 0, 0)$ & 2 & 0 & 0 & 0 & 0 & $2I$ \\
  $\tilde x_3=(0, 0, 1, 0, 0, 0)$ & 0 & 2 & 0 & 0 & 0 & $2J$ \\
  $\tilde x_4=(0, 0, 0, 1, 0, 0)$ & 0 & 0 & 1 & 0 & 0 & $K$ \\
  $\tilde x_5=(0, 0, 0, 0, 1, 0)$ & 0 & 1 & 1 & -1 & 0 & $J\,+\,K\,-\,L$ \\
  $\tilde x_6=(-3, -2, 0, 0, 0, 0)$ & 1 & -1 & -1 & -1 & 0 & $I\,-\,J\,-\,K\,-\,L$ \\
  $\tilde x_7=(6, 4, 1, 1, 1, 0)$ & 0 & 0 & -2 & 2 & 0 & $-2K\,+\,2L$ \\
  $\tilde x_8=(-6, -4, 0, -1, 0, 0)$ & 0 & -1 & 0 & 1 & 0 & $-J\,+\,L$ \\
  $x=(0, 0, 2, 1, 1, 3)$ & 0 & 2 & -2 & 2 & 2 & $2J\,-\,2K\,+\,2L\,+\,2M $ \\
  $y=(-3, -2, -2, -1, -1, -2)$ & 0 & 3 & -3 & 3 & 3 &$ 3J\,-\,3K\,+\,3L\,+\,3M$\\
  $z=(9, 6, 2, 1, 1, 0)$ & 0 & 0 & 0 & 0 & 1 & $M$ \\
  \hline 
  $\sum_i [D_i]$ & 6 & 6 & -6 & 6 & 6 & $6I\,+\,6J\,-\,6K\,+\,6L$ + 6M
  \end{tabular} 
  \label{FGLSMTable}
  \caption{GLSM Charges for $X_{\Sigma^{''}}$, the six-dimensional
    ambient toric variety for the elliptically fibered Calabi-Yau
    fourfold Y, which is a complete intersection of two
    hypersurfaces. We have indicated the generators of the fan.}
}

The reader should note, though, that the intermediate step of constructing the toric ambient space
$X_{\Sigma^{'}}$ of the base is not really necessary, since the GLSM
charges of homogeneous coordinates in $X_{\Sigma^{'}}$ are a subset of
the GLSM charges of homogeneous coordinates in $X_{\Sigma^{''}}$ and
one can easily deduce the charges of $y$ via the Calabi-Yau condition
and the adjunction formula. This yields
\begin{align}
c_1(T_Y) &= c_1(T_{X_{\Sigma^{''}}}) - N_{P_{X/\sigma}} - N_{P_{T^2}} \\ \notag &= \sum_i[D_i] - 6I - 2[D_y] = 6J - 6K + 6L + 6M - 2[D_y]=0,
\end{align}
with Poincar\' e duality implied. In the same way, one could determine
the charges of $x$, and again one could choose $z$ to only be charged
under $Q^5$. It is then possible to read off the class of $f$ and $g$,
or equivalently the Tate sections $a_n$, from the homogeneity of
$P_{T^2}$, without ever explicitly calculating the anticanonical
bundle. Of course, these different viewpoints are all closely tied
together, and the method one uses is a matter of preference.

\subsection{Instanton Zero Modes}

\label{sec:zero-modes}

So far, we have focused on discussing the F-theory model which will
have non-perturbative corrections to the $10\,10\,5_H$ Yukawa
coupling, but we have not yet discussed in detail the properties of
the instanton which generates the coupling. The reason for this is
that most known properties of euclidean branes in F-theory are known
only from the properties of the relevant instanton in the IIB model.

Charged zero modes in particular, which are the ones ultimately
responsible for generating the Yukawa coupling, are still poorly
understood from a purely F-theoretical point of view. What we have in mind
when making this statement is the description of F-theory as M-theory
with vanishing fiber. The properties of charged zero modes on the
euclidean M5 are not well understood. Nevertheless, F-theory is also
IIB at strong coupling, and in simple situations like ours the
description in terms of euclidean D3 branes is still expected to be
mostly correct. See, for example, \cite{Blumenhagen:2010ja} for a recent
paper which takes this viewpoint, obtaining a number of rules for the
spectrum of charged modes (these agree with the ones obtained in IIB,
except in the case where exceptional degenerations of the fiber
appear).

What this means, in practice, is that the known computation of the
superpotential coupling is isomorphic to the one done in IIB, except
for the issue of saturation of neutral zero modes ($\bar\tau$ modes in
particular). In this case there are more intrinsic ways of determining
this spectrum. The most well known way is using Witten's
characterization of fermionic zero modes as elements of the cohomology
of the structure sheaf of the divisor \cite{Witten:1996bn}. In the
rest of this section we will use this representation, together with
the result in \cite{Blumenhagen:2010ja} that the $\bar\tau$ mode can
be identified with an element of $H^{0,1}(\cD)$, to argue that the
$\bar\tau$ modes are projected out in our context. Before going into
that, we would also like to mention that one can also understand the
absence of dangerous neutral fermionic modes using the strongly
coupled IIB viewpoint \cite{Cvetic:2009ah}, so we already know what
the answer should be.\footnote{And since in this case we have a weakly
  coupled limit of the system, we also know the answer from a CFT
  analysis in IIB
  \cite{Argurio:2007qk,Argurio:2007vqa,Bianchi:2007wy,Ibanez:2007rs}.}
Nevertheless, computing the cohomology is an instructive exercise, to
which we now proceed.

\medskip

We study an $M5$ brane instanton on the fourfold divisor $\cD =
D_{X/\sigma} \cap D_{T^2} \cap D_5$, which is the intersection of the
base, fiber, and $D_5$ divisors in the ambient toric sixfold
$X_{\Sigma^{''}}$. The presence or absence of the
$\overline{\tau}_{\dot \alpha}$ zero modes for this instanton are
determined by the sheaf cohomology group
\begin{equation}
H^{1}(\cD,\cO_\cD),
\end{equation}
which can be related to sheaf cohomologies on $X_{\Sigma^{''}}$ via
Koszul sequences. For toric varieties, sheaf cohomology is equivalent
to the \v Cech cohomology groups $\check{H}^p(\cU,\cL)$, where $\cU$
is an open cover and $\cL$ is a line bundle on the toric
variety. Thus, our task is to compute $\check{H}^1(\cU, \cO_\cD)$ by
calculating the \v Cech cohomology groups of various line bundles on
$X_{\Sigma^{''}}$. We can do this easily using our implementation of
the algorithm in section~\ref{sec:cech}.

The divisor $\cD$ is the intersection of three divisors of the
sixfold, whose normal bundles are given by
\begin{align}
  \begin{split}
    N_{X/\sigma} &= \cO(6I) \\
    N_{T^2} & = \cO(6J-6K+6L+6M) \\
    N_{D_5} &= \cO(J+K-L).
  \end{split}
\end{align}
One can relate these objects on the ambient toric variety to the
structure sheaf on $\cD$ via the Koszul sequence
\begin{align}
0 \rightarrow \wedge^3 N^* \rightarrow \wedge^2 N^* 
\rightarrow N^* \rightarrow \cO_{X_{\Sigma^{''}}} \rightarrow \cO_{\cD} \rightarrow 0,
\end{align}
where $N^*$ is the dual of $N\equiv N_{X/\sigma} \oplus N_{T^2} \oplus
N_{D_5}$. For practical purposes, we split this into three short exact
sequences as
\begin{align}
0 & \rightarrow \wedge^3 N^* \rightarrow \wedge^2 N^* \rightarrow K_1 \rightarrow 0 \\
0 & \rightarrow K_1 \rightarrow N^* \rightarrow K_2 \rightarrow 0 \notag \\
0 & \rightarrow K_2 \rightarrow \cO_{X_{\Sigma^{''}}} \rightarrow \cO_{\cD} \rightarrow 0 \notag,
\end{align}
each of which gives a long exact sequence in cohomology, as outlined
in section~\ref{sec:cech}. Looking to the parts of the long exact
sequences relevant for the immediate calculation of
$H^{1}(\cD,\cO_\cD)$, we have
\begin{align}
\dots \rightarrow H^1&(X_{\Sigma^{''}},\cO_{X_{\Sigma^{''}}}) \rightarrow H^1(\cD,\cO_\cD)\rightarrow H^2(X_{\Sigma^{''}},K_2) \rightarrow H^2(X_{\Sigma^{''}},\cO_{X_{\Sigma^{''}}}) \rightarrow \dots \notag \\ \notag \\
&\dots \rightarrow H^2(X_{\Sigma^{''}},N^*) \rightarrow H^2(X_{\Sigma^{''}},K_2) \rightarrow H^3(X_{\Sigma^{''}},K_1) \rightarrow \dots \\ \notag \\
&\dots \rightarrow H^3(X_{\Sigma^{''}},\wedge^2N^*) \rightarrow H^3(X_{\Sigma^{''}},K_1) \rightarrow H^4(X_{\Sigma^{''}},\wedge^3N^*) \rightarrow \dots\,\,\, , \notag
\end{align}
one part for each short exact sequence. Calculating the cohomology of these line bundles on toric varieties, we arrive at the results
\begin{align}
H^4(X_{\Sigma^{''}},\wedge^3 &N^*)=0 \,\,\,\,\,\,\,\,\, H^3(X_{\Sigma^{''}},\wedge^2N^*)=0 \,\,\,\,\,\,\,\,\, H^2(X_{\Sigma^{''}},N^*)=0 \\
&H^1(X_{\Sigma^{''}},\cO_{X_{\Sigma^{''}}})=0 \,\,\,\,\,\,\,\,\, H^2(X_{\Sigma^{''}}, \notag\cO_{X_{\Sigma^{''}}})=0,
\end{align}
where Serre duality was useful for efficiently computing $H^4(X_{\Sigma^{''}},\wedge^3 N^*)$. Using these results, it is easy to see that
\begin{equation}
H^{1}(\cD,\cO_\cD)=0.
\end{equation}
We see that the $\overline{\tau}_{\dot \alpha}$ is projected out for
this instanton, as one might expect, since it is the lift of an $O(1)$
instanton in IIB. Thus, since the $\overline{\tau}_{\dot \alpha}$
modes are projected out and the cycle $\cD$ is rigid, we expect an
$M5$ brane instanton on $\cD$ to give a non-perturbative correction to
the $10\,10\, 5_H$ Yukawa coupling.

As a final word, there is a technical point that may be bothering the
reader: the instanton is on top of an $Sp(2)$ stack of branes, and
thus the fiber degenerates everywhere over its worldvolume. From this
point of view, computing the cohomology of the relevant divisor of the
fourfold seems to not be well-defined. Nevertheless, with the
definition that we have adopted here there are no issues, since
cohomologies of line bundles on the ambient toric space are always
well-defined. This point was further explored and reinforced in
\cite{Blumenhagen:2010ja}, where it was tested that the relevant
cohomology does not change under blow-ups of the geometry that smooth
out the degeneration of the fiber.

\subsection{The Tate Form for the Uplift}
\label{sec:tate}

In this section we discuss the details of the Tate form for the
elliptic fiber in the fourfold $Y$. We construct the most general form
of the sections $a_n$ of the Tate form in terms of the homogeneous
coordinates associated with the base and show that at a point in
complex structure moduli space the degenerations of the elliptic curve
recover two of the three gauge groups seen in the IIB limit. We show
that the third group $SO(6)$ is recovered only in Sen's weak coupling
limit. The results that we obtain in this example agree with, and
illustrate, the general discussion in section~\ref{sec:quantum-splitting}.

The fourfold $Y$, as mentioned, is an elliptic fibration over the base
$X/\sigma$. The elliptic fiber is often cast in the Weierstrass form
$y^2 = x^3 + fxz^4 + gz^6$, where $f\in
H^0(X/\sigma;K_{X/\sigma}^{-4})$ and $g\in
H^0(X/\sigma;K_{X/\sigma}^{-6})$ encode how the fiber varies over the
base. Often more useful in practice, however, is the Tate form
\begin{equation}
\label{eqn:Tate}
y^2+a_1\,xyz+a_3\,yz^3=x^3+a_2\,x^2z^2+a_4\,xz^4+a_6\,z^6,
\end{equation}
where $a_n\in H^0(X/\sigma;K_{X/\sigma}^{-n})$ instead encode the
variation of the fiber over the base. Particular combinations of the
$a_n$'s are grouped into variables
\begin{equation}
\label{eqn:Tatefg relations}
b_2=a_1^2 + 4\, a_2, \qquad b_4=a_1\, a_3 +2 a_4, \qquad b_6=a_3^2+4\, a_6
\end{equation}
which are related to $f$ and $g$ by
\begin{equation}
f=-{\textstyle \frac{1}{48}}( b_2^2 -24\, b_4), \qquad
  g=-{\textstyle \frac{1}{864}}( -b_2^3 + 36 b_2 b_4 -216 \, b_6).
\end{equation}
The discriminant, which encodes the locations of degeneration of the
elliptic fiber, and thus the 7-branes, takes the form
\begin{equation}
\Delta_F = 4f^3 + 27g^2 = -{\textstyle \frac{1}{4}}\, b_2^2\, (b_2 b_6- b_4^2) - 8 b_4^3 -27 b_6^2 +9 b_2 b_4 b_6 .
\end{equation}
The geometry of the base determines the explicit form of the sections
$a_n$ and the discriminant $\Delta_F$, from which the singularities of
the fiber, and thus the corresponding gauge groups of the 7-branes,
can be read off. The data relating the vanishing order of the sections
$a_n$ and the discriminant $\Delta_F$ to the singularity type is
reproduced in table~\ref{tab:TateTable}.

\TABLE{
        \centering
                \begin{tabular}{c|c|cc|cc@{${}\,\,\,\quad$}ccc}
                        sing.                        & discr.         & \multicolumn{2}{|c}{gauge enhancement}        & \multicolumn{5}{|c}{coefficient vanishing degrees}  \\
                        type                         & $\deg(\Delta)$ & type & group  & ${}\,\,\,\,\,\,a_1\quad{}$ & $a_2$ & $a_3$ & $a_4$ & $a_6{}_\big.$ \\ \hline\hline
                        $\mathrm{I}_0{}^\big.$               & 0      &            & ---          & 0 & 0 & 0     & 0     & 0 \\
                        $\mathrm{I}_1$                       & 1      &            & ---          & 0 & 0 & 1     & 1     & 1 \\
                        $\mathrm{I}_2$                       & 2      & $A_1$      & $SU(2)$      & 0 & 0 & 1     & 1     & 2 \\
                        $\mathrm{I}_3^{\,\mathrm{ns}}$       & 3      &            & [unconv.]    & 0 & 0 & 2     & 2     & 3 \\
                        $\mathrm{I}_3^{\,\mathrm{s}}$        & 3      &            & [unconv.]    & 0 & 1 & 1     & 2     & 3${}_\big.$ \\ \hline
                        $\mathrm{I}_{2k}^{\,\mathrm{ns}}$    & $2k$   & $C_{2k}$   & $SP(2k)$     & 0 & 0 & $k$   & $k$   & $2k$${}^\big.$ \\
                        $\mathrm{I}_{2k}^{\,\mathrm{s}}$     & $2k$   & $A_{2k-1}$ & $SU(2k)$     & 0 & 1 & $k$   & $k$   & $2k$ \\
                        $\mathrm{I}_{2k+1}^{\,\mathrm{ns}}$  & $2k+1$ &            & [unconv.]    & 0 & 0 & $k+1$ & $k+1$ & $2k+1$ \\
                        $\mathrm{I}_{2k+1}^{\,\mathrm{s}}$   & $2k+1$ & $A_{2k}$   & $SU(2k+1)$   & 0 & 1 & $k$   & $k+1$ & $2k+1$${}_\big.$ \\ \hline
                        $\mathrm{II}$                        & 2      &            & ---          & 1 & 1 & 1     & 1     & 1${}^\big.$ \\
                        $\mathrm{III}$                       & 3      & $A_1$      & $SU(2)$      & 1 & 1 & 1     & 1     & 2 \\
                        $\mathrm{IV}^{\,\mathrm{ns}}$        & 4      &            & [unconv.]    & 1 & 1 & 1     & 2     & 2 \\
                        $\mathrm{IV}^{\,\mathrm{s}}$         & 4      & $A_2$      & $SU(3)$      & 1 & 1 & 1     & 2     & 3 \\
                        $\mathrm{I}_0^{*\,\mathrm{ns}}$      & 6      & $G_2$      & $G_2$        & 1 & 1 & 2     & 2     & 3 \\
                        $\mathrm{I}_0^{*\,\mathrm{ss}}$      & 6      & $B_3$      & $SO(7)$      & 1 & 1 & 2     & 2     & 4 \\
                        $\mathrm{I}_0^{*\,\mathrm{s}}$       & 6      & $D_4$      & $SO(8)$      & 1 & 1 & 2     & 2     & 4 \\
                        $\mathrm{I}_1^{*\,\mathrm{ns}}$      & 7      & $B_4$      & $SO(9)$      & 1 & 1 & 2     & 3     & 4 \\
                        $\mathrm{I}_1^{*\,\mathrm{s}}$       & 7      & $D_5$      & $SO(10)$     & 1 & 1 & 2     & 3     & 5 \\
                        $\mathrm{I}_2^{*\,\mathrm{ns}}$      & 8      & $B_5$      & $SO(11)$     & 1 & 1 & 3     & 3     & 5 \\
                        $\mathrm{I}_2^{*\,\mathrm{s}}$       & 8      & $D_6$      & $SO(12)$     & 1 & 1 & 3     & 3     & 5${}_\big.$ \\ \hline
                        $\mathrm{I}_{2k-3}^{*\,\mathrm{ns}}$ & $2k+3$ & $B_{2k}$   & $SO(4k+1)$   & 1 & 1 & $k$   & $k+1$ & $2k$${}^\big.$ \\
                        $\mathrm{I}_{2k-3}^{*\,\mathrm{s}}$  & $2k+3$ & $D_{2k+1}$ & $SO(4k+2)$   & 1 & 1 & $k$   & $k+1$ & $2k+1$ \\
                        $\mathrm{I}_{2k-2}^{*\,\mathrm{ns}}$ & $2k+4$ & $B_{2k+1}$ & $SO(4k+3)$   & 1 & 1 & $k+1$ & $k+1$ & $2k+1$ \\
                        $\mathrm{I}_{2k-2}^{*\,\mathrm{s}}$  & $2k+4$ & $D_{2k+2}$ & $SO(4k+4)$   & 1 & 1 & $k+1$ & $k+1$ & $2k+1$${}_\big.$ \\ \hline
                        $\mathrm{IV}^{*\,\mathrm{ns}}$       & 8      & $F_4$      & $F_4$        & 1 & 2 & 2     & 3     & 4${}^\big.$ \\
                        $\mathrm{IV}^{*\,\mathrm{s}}$        & 8      & $E_6$      & $E_6$        & 1 & 2 & 2     & 3     & 5 \\
                        $\mathrm{III}^*$                     & 9      & $E_7$      & $E_7$        & 1 & 2 & 3     & 3     & 5 \\
                        $\mathrm{II}^*$                      & 10     & $E_8$      & $E_8$        & 1 & 2 & 3     & 4     & 5 \\
                        non-min                              & 12     &            & ---          & 1 & 2 & 3     & 4     & 6
                \end{tabular}
                \caption{Refined Kodaira classification resulting from
                  Tate's algorithm \cite{MR0393039}, from
                  \cite{Bershadsky:1996nh}. In order to distinguish
                  the ``semi-split'' case
                  $\mathrm{I}_{2k}^{*\,\mathrm{ss}}$ from the
                  ``split'' case $\mathrm{I}_{2k}^{*\,\mathrm{s}}$ one
                  has to work out a further factorization condition
                  which is part of the aforementioned algorithm, see
                  \S 3.1 of \cite{Bershadsky:1996nh}.}
                \label{tab:TateTable}
}

In our case we have $a_n \in H^0(X/\sigma;\mathcal O(n(J-K+L))$. Since
they must be global sections, the orders of vanishing of the
homogeneous coordinates $x_i\in X_{\Sigma^{'}}$ appearing in the
monomials must be positive. Thus, the divisors corresponding to the
monomials must be \emph{effective}:
\begin{equation}
[D] = \sum_i n_i [D_i] = n(J-K+L) \,\,\,\,\,\,\,\, n_i\ge 0 \,\,\,\,\,\,\,\, \forall n_i.
\end{equation}
Satisfying this condition for the case at hand yields the result
\begin{equation}
n_3 = -n_7 + n \,\,\,\,\,\,\,\,\,\,\,\,\,\,\, n_5=2n_7-n \,\,\,\,\,\,\,\,\,\,\,\,\,\,\, n_1=n_2=n_4=n_6=n_8=0,
\end{equation}
which completely determines the allowed monomials in each section $a_n$. Note that this gives
\begin{equation}
a_1 = c_0\tilde x_5\tilde x_7 \,\,\,\,\,\,\,\,\,\, a_2=c_1\tilde x_3\tilde x_7+c_2\tilde x_5^2\tilde x_7^2,
\end{equation}
which leads to $b_2=c_0^2\tilde x_5^2\tilde x_7^2+4c_1\tilde x_3\tilde
x_7+4c_2\tilde x_5^2\tilde x_7^2$. It can be shown that in Sen's
limit, the orientifold is located at $b_2=0$, which in our case
corresponds to O7-planes on the divisor $[O7]=[D_3]+[D_7]$. This is
precisely the result of the simple analysis on the IIB side.
Continuing this analysis for the sake of examining possible gauge
enhancements gives
\begin{align}
  a_3=c_3\tilde x_5^3\tilde x_7^3 + c_4\tilde x_3\tilde x_5\tilde
  x_7^2 \,\,\,\,\,\,\,\,\,\,\,\, a_4=c_5\tilde x_5^4\tilde x_7^4 +
  c_6\tilde x_3\tilde x_5^2\tilde x_7^3 + c_7\tilde x_3^2\tilde x_7^2
  \\ \notag a_6=c_8\tilde x_5^6\tilde x_7^6 + c_9\tilde x_3\tilde
  x_5^4\tilde x_7^5 + c_{10}\tilde x_3^2\tilde x_5^2\tilde x_7^4 +
  c_{11}\tilde x_3^3\tilde x_7^3. \,\,\,\,\,\,\,\,\,\,\,\,\,\,\,\,\,\,
\end{align}
This is the most general form for the Tate sections in this model,
from which the ``minimal'' gauge enhancements can be read off. For
example, using table~\ref{tab:TateTable}, it can be
seen from the order of vanishing along $D_7$ that it has minimal gauge
group $G_2$ (a similar phenomenon was found in
\cite{Blumenhagen:2009qh}). Rather than constructing the most general
allowed fibration for this model, however, we would like to reproduce
as much of the IIB physics as possible in the F-theory lift. Moving to
a point in complex structure moduli space where
\begin{align}
  a_2=c_1\tilde x_3\tilde x_7 \,\,\,\,\,\,\,\,\,\, a_3=4c_1c_3\tilde x_3\tilde x_5\tilde x_7^2 \,\,\,\,\,\,\,\,\,\,  a_1=a_4=a_6=0 \\
  \Delta_F = -256c_1^4c_3^2\tilde x_3^4\tilde x_5^2\tilde
  x_7^7(27c_3^2\tilde x_5^2\tilde x_7 + c_1\tilde x_3), \notag
  \,\,\,\,\,\,\,\,\,\,\,\,\,\,
\end{align}
it is readily seen that the gauge groups along $D_7$ and $D_5$ are
$SO(10)$ and $Sp(2)$, respectively, as is the case in IIB. However,
recovering the factor of $SO(6)$ along $D_3$ requires taking $c_3 \to
0$, which sends $\Delta_F \to 0$ everywhere and thus $g_s \to 0$. This
is precisely Sen's limit \cite{Sen:1997gv}. This agrees beautifully
with the discussion in section~\ref{sec:quantum-splitting}.

\section{Conclusions}
In this paper we have addressed a number of conceptual and technical
issues which arise in the analysis of instanton effects in
F-theory.

\medskip

We started in section \ref{sec:quantum-splitting} by explaining the
reason for an obstruction that can appear when trying to lift certain
stacks of branes in IIB to F-theory. This also allowed us to make
predictions about which IIB brane configurations are obstructed. In
the process, we described in some detail the behavior of D7 branes as
we go from large to vanishing flavor masses.

\medskip

We continued in section~\ref{sec:cech} by discussing a way of
computing sheaf bundle cohomology on toric varieties. This is
essential for showing that the $\overline{\tau}_{\dot \alpha}$ mode is
projected out, which requires calculation of the cohomology group
$H^{1}(\cD,\cO_\cD)$, where $\cD$ is the fourfold divisor which the
$M5$ wraps. One can compute this sheaf cohomology by calculating \v
Cech cohomology of line bundles on the ambient toric variety, and
running it through the long exact sequences in cohomology given by the
splits of the Koszul sequence. Specifically, we review how to compute
\v Cech cohomology on toric varieties in general in
section~\ref{sec:cech-general} and give an illustrative example on
$dP_1$ in section~\ref{sec:dp1-cech}. In \ref{sec:koszul}, we discuss
the Koszul sequence, which gives a long exact sequence in cohomology
which allows one to compute the cohomology $H^{1}(\cD,\cO_\cD)$ by
knowing information about \v Cech cohomology of line bundles on the
ambient toric variety. Finally, in section~\ref{sec:computer
  implementation}, we provide some details about where to find our
ready-to-use computer implementation of the algorithm.

\medskip

In section~\ref{sec:example} we illustrated the considerations in the
previous sections in a particular example. We introduced in section
\ref{sec:IIB} the Calabi-Yau threefold $M_2^{(dP_9)^2}$, henceforth
called $X$, as a Calabi-Yau hypersurface in a four-dimensional toric
variety $X_\Sigma$. In section \ref{sec:fourfold}, we performed the
F-theory lift of the IIB orientifold compactification on $X$,
following \cite{Collinucci:2008zs}. For the sake of clarity, we
presented the lift in two steps. First, we presented the fourfold base
$X/\sigma$ as a hypersurface in a four-dimensional toric variety
$X_{\Sigma^{'}}$ by properly modding out by the orientifold action
$\sigma$. Next, we presented the uplifted Calabi-Yau fourfold $Y$ as a
complete intersection of two hypersurfaces in a six-dimensional toric
variety $X_{\Sigma^{''}}$, one associated to the fiber and one to the
base.

In section \ref{sec:zero-modes}, we addressed the issue of instanton
zero modes in the F-theory uplift of the IIB orientifold
compactification on $X$. In the F-theory lift, we showed the absence
of the fermionic $\overline{\tau}_{\dot \alpha}$ zero modes for a
vertical $M5$ brane which is necessary for the generation of the
$10\,10\,5_H$ Yukawa coupling.

In section~\ref{sec:tate} we determined explicitly the Tate sections
and discriminant for the F-theory lift, allowing us to see the
location of seven branes as divisors in the base over which the fiber
degenerates, as well as their associated gauge group. At a generic
point in moduli space, this data determines the ``minimal'' gauge
enhancements along the seven branes, but we showed a point in complex
structure moduli space which recovers, in F-theory, the proper
location of the orientifold and two of the three gauge seven branes
seen on the IIB side. Interestingly, it is only in Sen's IIB limit
that the proper enhancement of the third gauge seven brane is
obtained, in agreement with the general discussion of
section~\ref{sec:quantum-splitting}.

\medskip

F-theory compactifications provide a rich field of study both for
formal and phenomenological questions. The way the results in this
paper came to be nicely illustrates this connection: we set out to
study a particular model with some nice phenomenological features, and
we were driven to fascinating questions in Seiberg-Witten theory and
algebraic geometry. There is no doubt that there are still plenty of
interesting phenomena to be elucidated in the quest for fully
realistic F-theory models.

\acknowledgments

We would like to acknowledge interesting discussions with Lara
Anderson, Ralph Blumenhagen, B.G. Chen, Andres Collinucci, Ron Donagi,
Josh Guffin, Benjamin Jurke, Jeffrey C.Y. Teo and Timo Weigand. We
would also like to thank the authors of \cite{Blumenhagen:2010pv} for
informing us of their upcoming work prior to publication. We are also
grateful to the editor of JHEP, who suggested a rearrangement of the
sections to increase the clarity of the exposition. We gratefully
acknowledge the hospitality of the KITP during the Strings at the LHC
and in the Early Universe program for providing a stimulating
environment during the completion of this work. I.G.E. thanks
N. Hasegawa for kind support and constant encouragement. This research
was supported in part by the National Science Foundation under Grant
No. NSF PHY05-51164, DOE under grant DE-FG05-95ER40893-A020, NSF RTG
grant DMS-0636606 and Fay R. and Eugene L. Langberg Chair.

\bibliographystyle{JHEP}
\bibliography{refs}

\end{document}